\documentclass[acmsmall,screen,nonacm]{acmart}

\renewcommand\footnotetextcopyrightpermission[1]{}
\usepackage{booktabs}
\usepackage{multirow}
\usepackage{array}
\usepackage{tabularx}
\usepackage{adjustbox}
\usepackage{graphicx}
\usepackage{xcolor}
\usepackage{colortbl}
\usepackage{listings}
\usepackage{tcolorbox}
\usepackage{enumitem}

\tcbuselibrary{skins,breakable}

\definecolor{pgMethod}{HTML}{1B528C}      % ShellPlan
\definecolor{pgMethodTint}{HTML}{DBE3ED}
\definecolor{pgAlert}{HTML}{B45540}       % Violations and out-of-policy effects
\definecolor{pgAlertTint}{HTML}{F3E4E0}
\definecolor{pgBase}{HTML}{8C959D}        % Baselines and neutral borders
\definecolor{pgInk}{HTML}{2B323A}         % Text ink
\definecolor{pgInkTint}{HTML}{F4F5F6}
\definecolor{boxbg}{named}{pgInkTint}
\definecolor{boxborder}{named}{pgBase}
\definecolor{codebg}{named}{pgInkTint}
\AtBeginDocument{\hypersetup{linkcolor=pgMethod,citecolor=pgMethod,urlcolor=pgMethod}}

\newtcolorbox{findingbox}[1]{%
  enhanced,
  colback=boxbg,
  colframe=boxborder,
  boxrule=0.75pt,
  arc=2pt,
  left=5pt, right=5pt, top=3pt, bottom=3pt,
  title={\textbf{#1}},
  coltitle=pgInk,
  colbacktitle=boxbg,
  titlerule=0pt,
  fonttitle=\footnotesize\sffamily,
  fontupper=\footnotesize
}

\newtcolorbox{disclosurebox}{%
  enhanced,
  colback=pgAlertTint,
  colframe=pgAlert,
  boxrule=0.7pt,
  arc=2pt,
  left=6pt, right=6pt, top=4pt, bottom=4pt,
  fontupper=\footnotesize
}

\newcommand{\code}[1]{\texttt{\small #1}}
\newcommand{\parsegap}{\textsc{ParseGap}}
\newcommand{\shellplan}{\textsc{ShellPlan}}

\begin{document}

\title{\parsegap: Testing What Coding Agents Authorize and What Shells Execute}

\author{Jieyu Li}
\affiliation{%
  \institution{College of Computer Science and Artificial Intelligence, Fudan University}
  \city{Shanghai}
  \country{China}
}
\email{26113050216@m.fudan.edu.cn}

\begin{abstract}
LLM coding agents decide whether to execute a shell command by parsing the command string against a policy and then passing the same raw string to a system shell. The approval gate therefore adjudicates its own reading of the string, while the shell executes its own. We present \parsegap, a differential testing framework that evaluates each input through an unmodified vendor checker and through GNU Bash~5.2 in an isolated container, flagging a violation whenever a checker permits a command whose observed runtime effects violate the policy. Across eight production coding agents, \parsegap{} catalogs 17 deduplicated root causes and pinpoints three boundaries where authorization and execution diverge: at the parse, where an upstream \code{tree-sitter-bash} defect hides a second command from Mistral Vibe and Gemini CLI; at the utility, where every checker permits read-only inspection binaries whose arguments write files or launch programs (9--47 of 55 cases); and at confinement, where matching an allow rule in OpenAI Codex CLI also disables its read-only OS sandbox. Live model-driven trials across two frontier models produced targeted policy-violating effects in all 16 agent--model pairs, including 14 under configured restriction rules.

To eliminate the second reading, we design \shellplan, which compiles a strict Bash subset once into a typed execution plan, validates utility arguments against positive capability specifications, and dispatches the approved plan directly via \code{execve} without invoking a shell. Over a formal operational model of command gates, we prove single-interpretation dispatch and bound the programs and files touched by admitted plans of specified utilities under explicit, empirically audited assumptions. \shellplan{} admitted 0/4,174 active attack cases across six static suites and achieved 0/90 attack hits in a ten-model controlled agent benchmark (62/90 unguarded) at a median gate latency of 1.28\,ms per task. It preserved 100\% benign completion on standard development tasks across six frontier models (36/36; 56/60 overall including prompted syntax-rewrite tasks) and completed 74/100 benign tasks across all ten models (93/100 unguarded), with escalation losses concentrated in smaller models. A pre-registered external evaluation on 60 real agent trajectories and 1,523 NL2Bash commands confirms 0/348 over-allows outside modeled exceptions and 343/343 byte-exact Bash matches on SWE-bench Verified, while delimiting remaining scope boundaries (stateful \code{cd} chains in OpenHands, one workspace-write metadata edge case, and two fidelity mismatches).
\end{abstract}


\maketitle

\begin{disclosurebox}
\textbf{Coordinated Disclosure Notice (arXiv v1 Preprint).} All newly discovered enforcement defects reported in this work were privately disclosed to the affected vendors and upstream parser maintainers on November~22,~2026 under a standard 90-day coordinated disclosure window. To protect users during remediation, this preprint redacts exact trigger byte sequences, internal source line pointers, and step-by-step reproduction recipes for unpatched issues; full unredacted witnesses and archival artifacts will be restored in v2 upon patch availability.
\end{disclosurebox}

% ============================================================================
\section{Introduction}
\label{sec:intro}
% ============================================================================

Coding agents execute shell commands directly inside developer workspaces~\cite{yang2024sweagent,wang2025openhands}. Because untrusted repository files, build logs, and issue trackers can steer the commands an agent proposes~\cite{greshake2023indirect,zhan2024injecagent,liu2025aishelljack}, coding agents place a command approval gate in front of shell execution to auto-approve safe commands and \emph{escalate} risky ones to the developer. Our inventory of public security advisories from July~2025 through October~2026 (deduplicated CVE/GHSA identifiers) counts 171 shell-approval-bypass advisories: 58 across ten mainstream coding agents (Claude Code, Cursor, Roo Code, Mistral Vibe, Codex CLI, Gemini CLI, Copilot CLI, GitHub Copilot, Amazon Q, and Ollama) and 113 in OpenClaw, an open-source agent exposing the same shell-gating surface.

\begin{figure}[!t]
  \centering
  \includegraphics[width=\textwidth]{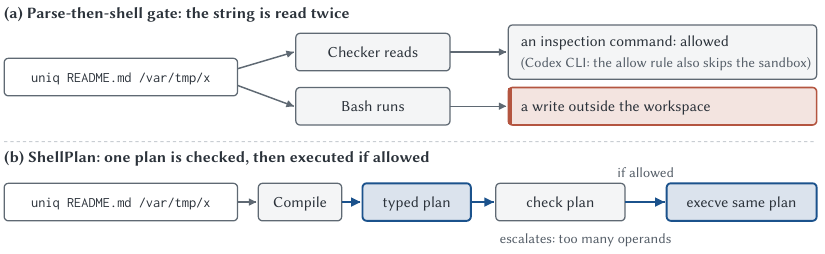}
  \Description{Comparison of a parse-then-shell gate and ShellPlan on the running example uniq README.md /var/tmp/x.}
  \caption{The running example under two gate architectures. (a)~A parse-then-shell gate approves the raw string as a read-only inspection command, while Bash executes it as a file write. (b)~\shellplan{} compiles the string once into a typed plan, validates utility arguments against positive specifications, and dispatches that same plan via \code{execve} only if allowed.}
  \label{fig:paths}
\end{figure}

Consider a read-only policy that allows standard inspection utilities such as \code{uniq}, and the command \code{uniq README.md /var/tmp/x}. Six of the eight production agents we study allow it. Bash then invokes \code{uniq}, which treats its second positional operand as an output file and writes outside the workspace; in Codex CLI, matching the allow rule additionally disables the \code{bwrap}+Landlock sandbox that would otherwise block the write (Figure~\ref{fig:paths}a). Each component behaves according to its local specification: the checker matched an allowed utility name, the shell executed the raw string, and \code{uniq} honored its POSIX interface. The security failure emerges strictly between the layers.

All eight production coding agents share this dual-reading architecture. A checker parses the string with its own parser (\code{shell-quote} tokenizers, \code{tree-sitter-bash} ASTs, or Starlark prefix rules) and renders a decision on the commands it extracts. The shell parses the raw string a second time, each invoked utility interprets its own arguments and environment, and in Codex CLI the approval decision also toggles OS confinement. Authorization and execution therefore diverge at three distinct boundaries: \emph{the parse}, \emph{utility argument semantics}, and \emph{confinement}. Prior studies measure checker verdicts, end-to-end exploit rates, or harness mechanisms~\cite{chen2026cmdneedle,jing2026policygate,zhu2026harnesssecbench}. We pair each unmodified checker's verdict with the actual runtime effects of executing the same input under GNU Bash~5.2 and classify every violation by boundary (Definition~\ref{def:divergence}). Our core insight is that a gate's decision governs only its own reading of a command and remains sound only when execution is bound directly to that validated representation.

We first measure where production gates fail. \parsegap{} runs every input through an unmodified vendor checker and through GNU Bash~5.2 inside an isolated Linux container, traces launched processes and persistent filesystem mutations, and reports a violation whenever the checker allows an input whose runtime effects the reference policy prohibits (Sections~\ref{sec:background}--\ref{sec:problem}). This answers \textbf{RQ1 (Divergence):} \emph{at which boundaries do production checkers allow policy-violating inputs, and does OS sandboxing still contain them?} and \textbf{RQ2 (Consequence):} \emph{do these divergences yield targeted effects when frontier models drive unmodified agent CLIs?}

We then eliminate the second reading. \shellplan{} compiles a strict Bash subset once into a typed execution plan, checks utility arguments against positive capability specifications, validates Git repository configurations, and executes admitted plans directly via \code{execve} without invoking a shell or altering OS confinement (Figure~\ref{fig:paths}b). Over a formal operational model, we prove that \shellplan{} executes strictly the plan it authorizes (Theorem~\ref{thm:single}) and bound the processes and files touched by admitted plans of specified utilities under explicit assumptions (Theorem~\ref{thm:contain}). We evaluate \shellplan{} through \textbf{RQ3 (Security and Cost):} \emph{under matched policies, what attacks does single-interpretation execution block, and what legitimate workflows does it escalate?} and \textbf{RQ4 (Empirical Boundaries):} \emph{where do \shellplan's guarantees stop holding across white-box audits, real-world agent trajectories, and deployment inside an unmodified vendor CLI?}

Across all eight production agents, every checker allowed utility-mediated policy violations, and live model-driven trials produced targeted effects in all 16 agent--model pairs (14 under configured restriction rules). \shellplan{} admitted 0/4,174 active attack cases across six static suites and achieved 0/90 attack hits across ten models at 1.28\,ms median gate latency per task, preserving 100\% completion on standard development tasks across all six frontier models (36/36; 74/100 overall including smaller models and prompted syntax-rewrite tasks, vs.\ 93/100 unguarded).

\textbf{Contributions.} This paper makes four contributions:
\begin{enumerate}[leftmargin=1.5em,itemsep=1.5pt]
  \item \textbf{Formalization and Differential Oracle:} We formalize parse-then-shell gates and build \parsegap, a differential testing framework that scores unmodified vendor checkers against observed Bash~5.2 runtime effects (Sections~\ref{sec:problem}--\ref{sec:methodology}).
  \item \textbf{Empirical Study Across Eight Agents:} We catalog 17 deduplicated root causes (\code{RC-01}--\code{RC-17}), including Codex CLI's policy--sandbox inversion (\code{RC-14}), Qwen Code's permission-precedence override (\code{RC-16}), an upstream \code{tree-sitter-bash} cross-product defect (\code{RC-17}), and an unsandboxed startup repository-filter execution in Codex CLI (Section~\ref{sec:empirical}).
  \item \textbf{Compile-Once, Direct-\code{execve} Gate:} We design \shellplan, which checks and directly executes a single compiled execution plan with positive utility specifications and open-time redirection hardening, accompanied by formal guarantees (Section~\ref{sec:design}).
  \item \textbf{Comprehensive Evaluation and Real-CLI Integration:} We evaluate \shellplan{} across 4,174 active attacks, a 760-run ten-model agent benchmark, three white-box audit rounds, a pre-registered external study of 60 SWE-bench/OpenHands trajectories and 1,523 NL2Bash commands, and live integration inside unmodified Codex CLI (Section~\ref{sec:eval}).
\end{enumerate}

% ============================================================================
\section{Background, Threat Model, and Running Example}
\label{sec:background}
% ============================================================================

\subsection{Studied Coding Agents and Execution Architecture}

Table~\ref{tab:products} summarizes the eight studied production coding agents, spanning diverse parsing strategies, permission engines, and confinement layers. Across all tested execution modes, every product hands approved command text to a system shell: Codex, Gemini, Qwen, Mistral Vibe, OpenCode, Cline, and Continue spawn \code{/bin/bash} or \code{/bin/sh} as a child process, while Roo Code dispatches the command string inside its terminal shell.

\begin{table*}[!t]
\centering
\caption{Eight studied production coding-agent implementations (pinned November~2026). All eight pass approved command strings to a system shell in the tested execution modes.}
\label{tab:products}
\scriptsize
\setlength{\tabcolsep}{3.5pt}
\renewcommand{\arraystretch}{1.06}
\begin{tabularx}{\textwidth}{@{}llllX@{}}
\toprule
\textbf{Product} & \textbf{Checker (isolated)} & \textbf{CLI (live)} & \textbf{Command Parser} & \textbf{Policy Engine \& Confinement} \\
\midrule
Roo Code & \code{v3.54.0} & CLI \code{0.1.17}$^{\dagger}$ & \code{shell-quote} + rewrites & Prefix allow/deny list; no built-in OS sandbox \\
Cline & \code{main} (2026-11-12) & CLI \code{3.0.74} & Regex / token guard & Plan-mode RO guard; Act mode has no allowlist \\
Continue & \code{main} (2026-11-12) & CLI \code{1.5.47} & Hand-written AST & Built-in security classifier; headless \code{Bash=allow} \\
OpenCode & \code{dev} (2026-11-12) & CLI \code{2.3.0} & \code{tree-sitter-bash} & Glob allow/deny rules; headless \code{run} auto-allows \code{ask} \\
Gemini CLI & \code{0.70.0} & CLI \code{0.70.0} & WASM \code{tree-sitter-bash} & Prioritized TOML \code{PolicyEngine} (\code{allow}/\code{deny}) \\
Qwen Code & \code{0.28.3} & CLI \code{0.28.3} & \code{tree-sitter-bash} & \code{PermissionManager} + substitution check + virtual edit deny \\
Mistral Vibe & \code{3.1.4} & CLI \code{3.1.4} & Py \code{tree-sitter-bash} 0.25.1 & Prefix allow/deny + 15 per-binary guardrails + path check \\
OpenAI Codex & \code{rust-v0.171.0} & CLI \code{0.171.0} & Rust \code{tree-sitter} & Starlark \code{execpolicy} + danger classifier + Linux \code{bwrap}+Landlock \\
\bottomrule
\end{tabularx}
\begin{flushleft}
\scriptsize
$^{\dagger}$Roo Code CLI \code{0.1.17} executes unmodified upstream \code{extension.js} (SHA-256 \code{5423e729...}) inside a headless config-injection wrapper \code{lib/index.js}.
\end{flushleft}
\end{table*}

\subsection{Threat Model and Policy Distinction}

An agent operates inside a developer workspace $W$ (a Git repository) and ingests untrusted repository artifacts before proposing shell commands~\cite{greshake2023indirect,liu2025aishelljack}. The adversary controls untrusted text read by the agent as well as files, symbolic links, Git metadata, and local scripts inside $W$; inherited environment variables, host configuration outside $W$, and system executables outside $W$ are trusted. Confidentiality (reading files outside $W$ via an authorized command) and denial of service are out of scope.

Every test run is scored against a unified \emph{reference effect policy} $\Pi_{\text{ref}}$. Each product's \emph{configured policy} $\Pi_{\text{prod}}$ expresses the same allow and deny entries as faithfully as that vendor's policy syntax allows. Several CLI modes permit targeted actions by documented default (Continue headless \code{Bash=allow}, Cline Act mode, and Codex \code{workspace-write} for ordinary workspace writes); we classify these explicitly as \emph{permissive modes} and report them apart from policy-rule bypasses.

\subsection{Running Example: Policy--Sandbox Inversion in Codex CLI (\code{RC-14})}
\label{sec:motivating}

Suppose a developer launches Codex CLI~\code{0.171.0} in read-only sandbox mode (\code{--sandbox read-only}), configures Starlark prefix allow rules for standard inspection binaries such as \code{uniq} and \code{sed}, explicitly forbids \code{touch}, and the agent proposes \code{uniq README.md /var/tmp/x}. Codex already restricts its parsed subset to plain word-only commands joined by \code{\&\&}, \code{||}, \code{;}, and \code{|}. Because \code{uniq README.md /var/tmp/x} belongs to this plain subset, the prefix rule engine matches the leading command name \code{uniq}, ignores the second positional output operand, and returns \code{allow}. Crucially, when every sub-command matches an allow rule, Codex's approval layer sets an internal sandbox-bypass flag, causing the tool orchestrator to dispatch the command with no OS sandbox (\code{bwrap} and Landlock skipped entirely) despite \code{--sandbox read-only}.

Under \code{strace -f}, live Codex CLI sessions driven by \code{gpt-5.6-luna} and \code{deepseek-v4.1-flash} spawned zero \code{bwrap} processes on all allowlisted witnesses: \code{uniq} created an external file outside $W$, and an allowlisted \code{sed} invocation with an embedded execute directive spawned \code{/bin/sh} to run the explicitly forbidden \code{touch}. By contrast, running the same utility-semantics (55 cases) and workspace-integrity (28 cases) suites inside Codex's isolated read-only \code{bwrap} sandbox without prefix allow rules yielded zero persistent filesystem mutations (0/55 and 0/28). Because approval and confinement share a single decision, configuring read-only allow rules silently removed OS confinement.

% ============================================================================
\section{Problem Formulation}
\label{sec:problem}
% ============================================================================

We formalize the oracle behind \parsegap{} and the safety guarantees of \shellplan.

\textbf{Commands and Runtime Effects.} A command string $s$ executes in a state $\sigma = (W, F_{\mathrm{ext}}, E)$ comprising workspace $W$, external filesystem state $F_{\mathrm{ext}}$, and inherited environment $E$. Bash's grammar decomposes $s$ into simple commands $\mathit{cmds}_B(s)=\langle c_1,\dots,c_n\rangle$, each a sequence of unexpanded words; expansion and control flow then depend on $\sigma$. Let $\mathit{Eff}(s,\sigma)$ denote the effects observed when unconfined Bash executes $s$ in $\sigma$: successfully launched processes (executable path and argument vector) and persistent mutations across monitored paths. A reference policy $\Pi$ partitions effects into permitted and prohibited sets; under read-only profile \code{P1} (Section~\ref{sec:methodology}), every persistent filesystem mutation and every process launch outside the allowlist is prohibited.

\begin{definition}[Parse-then-Shell Gate]
\label{def:gate}
A gate $G = (\mathit{parse}_G, \mathit{rules}_G, \kappa_G)$ parses $s$ into simple commands via $\mathit{parse}_G$, decides $\delta_G(s)=\mathit{rules}_G(\mathit{parse}_G(s))\in\{\textsf{allow},\textsf{ask},\textsf{deny}\}$, and on \textsf{allow} dispatches the raw string $s$ to a system shell under a confinement configuration $\kappa_G(s)$ that the gate may select.
\end{definition}

\begin{definition}[Active Input and Policy Violation]
\label{def:violation}
An input $s$ is \emph{active} under $(\Pi,\sigma)$ if $\mathit{Eff}(s,\sigma)$ contains at least one $\Pi$-prohibited effect. A gate $G$ commits a \emph{violation} on $s$ iff $s$ is active and $\delta_G(s)=\textsf{allow}$.
\end{definition}

Evaluating Definition~\ref{def:violation} against observed Bash effects guarantees that malformed or inert strings never inflate or dilute security denominators.

\begin{definition}[Divergence Boundaries]
\label{def:divergence}
A violation on $s$ carries one or more of three boundary labels:
\textbf{(D1) Parse divergence:} $\mathit{parse}_G(s)$ and $\mathit{cmds}_B(s)$ disagree on a command boundary or word list, and that disagreement enables $\delta_G(s)=\textsf{allow}$.
\textbf{(D2) Utility divergence:} the command decompositions agree, and the prohibited effect arises from a utility's arguments, inline environment assignments, or redirections.
\textbf{(D3) Confinement divergence:} executing $s$ under $\kappa_G(s)$ produces a prohibited effect that executing $s$ under the user-configured sandbox mode prevents in the same state $\sigma$.
\end{definition}

Rule-precedence faults, permissive headless defaults, and danger-classifier gaps fall outside D1--D3 and are attributed separately. In the running example, the parse agrees while \code{uniq} writes its second operand (D2), and Codex's allow decision disables the read-only sandbox that blocks the write (D3). In \code{RC-17}, the checker sees a single \code{echo} while Bash executes \code{echo} followed by a second command (D1).

% ============================================================================
\section{\parsegap: Differential Testing and Evaluation Methodology}
\label{sec:methodology}
% ============================================================================

\parsegap{} operationalizes Definition~\ref{def:violation} across four complementary evaluation stages: (1)~\emph{Component Evaluation} measuring unmodified vendor checkers, isolated OS sandboxes, and \shellplan{} on unified corpora; (2)~\emph{Entrypoint Confirmation} verifying CLI execution under fixed mock model outputs; (3)~\emph{Live Product Trials} driving unmodified vendor CLIs with frontier LLMs; and (4)~\emph{Controlled Agent Benchmarks} comparing defenses head-to-head across ten models. Every system is scored along three dimensions with explicit denominators: \emph{Security} (fraction of active inputs allowed by a checker or producing monitored prohibited effects under live/sandboxed execution), \emph{Utility} (fraction of eligible read-only commands escalated, paired with the count of non-read-only commands over-allowed on the same corpus), and \emph{Fidelity} (agreement of exit code, \code{stdout}, and \code{stderr} with GNU Bash~5.2 in identical state, reporting exact matches, clean refusals, and mismatches separately).

\begin{figure}[!t]
  \centering
  \includegraphics[width=\textwidth]{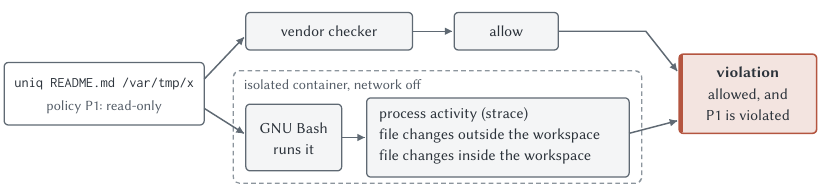}
  \Description{Differential testing workflow of ParseGap comparing an unmodified vendor checker against GNU Bash 5.2 runtime effects.}
  \caption{One differential test in \parsegap. Each command string is evaluated by an unmodified vendor checker and executed by GNU Bash~5.2 in an isolated container; a violation occurs when an active prohibited effect is permitted (Definition~\ref{def:violation}).}
  \label{fig:oracle}
\end{figure}

\textbf{Reference Execution and Effect Monitors.} Figure~\ref{fig:oracle} illustrates one differential test. Inside an isolated Linux container (\code{--network none}, GNU Bash~5.2, workspace \code{/work} reset before every run), \parsegap{} invokes the vendor's unmodified checker binary or module without reimplementing vendor logic, and executes $s$ separately via \code{/bin/bash -c}. Three monitors capture $\mathit{Eff}(s,\sigma)$: (1)~completed \code{execve}/\code{execveat} calls via \code{strace -f} (excluding the outer launcher); (2)~persistent diffs on monitored external paths (\code{/var/tmp} and \code{/tmp/canary}); and (3)~SHA-256 content and existence diffs inside \code{/work}, separating \code{.git/*} metadata from ordinary workspace files. Isolated \code{codex sandbox} evaluations (\code{bwrap -P ro}, \code{-P ww}) score persistent filesystem monitors~(2) and~(3).

\textbf{Three Policy Profiles.} (1)~\textbf{\code{P1} (Read-Only)} allowlists 18 inspection binaries (\code{ls}, \code{cat}, \code{head}, \code{tail}, \code{wc}, \code{echo}, \code{pwd}, \code{which}, \code{file}, \code{stat}, \code{du}, \code{df}, \code{rg}, \code{grep}, \code{sort}, \code{uniq}, \code{jq}, \code{printenv}), 6 \code{git} read-only subcommands (\code{status}, \code{diff}, \code{log}, \code{show}, \code{branch}, \code{remote -v}), and constrained \code{find}/\code{sed}, without a denylist. (2)~\textbf{\code{P2} (Dev-Native)} combines a narrower 11-entry allowlist (\code{echo}, \code{ls}, \code{cat}, \code{git status/diff/log}, \code{rg}, \code{grep}, \code{find}, \code{sed}, \code{npm test}) with explicit deny rules (\code{rm}, \code{curl}, \code{sudo}, \code{chmod}, \code{chown}, plus \code{touch} in OpenCode), so allow rates can decrease from \code{P1} to \code{P2}. (3)~\textbf{\code{P3} (Denylist)} pairs a broad allowlist (18 utilities, or \code{*} in OpenCode) with deny rules (\code{rm}, \code{sudo}, \code{curl}, \code{wget}, \code{touch}, \code{chmod}, \code{chown}, \code{sh}, \code{bash}, \code{dd}). Active membership is fixed once under \code{P1} Bash~5.2 execution; \code{P2}/\code{P3} columns in Table~\ref{tab:t1} measure allow rates on that exact \code{P1}-active set.

\textbf{Challenge and Utility Suites.} Six attack suites yield \textbf{4,174 active cases} (counted per suite without cross-suite deduplication): \textbf{CVE-Replay} (33/34 advisory PoCs), \textbf{SyntaxMatrix} (405/420 grammar combinations), \textbf{Attack-1K} (968/1,007 adversarial templates), \textbf{Fuzz-3K} (2,685/3,000 differential mutations), \textbf{Attack-BinSem} (55/57 man-page utility-semantics cases, including 18 execution cases, plus 5 benign controls of which 4 are spec-covered), and \textbf{Workspace-Integrity} (28/28: 8 \code{.git/*} metadata targets and 20 ordinary workspace files). Two benign utility suites measure escalation and over-allow: \textbf{Benign-Dev} (160 read-only and 40 non-read-only commands) and \textbf{Benign-Ext} (184 read-only candidates and 316 non-read-only commands). Filtering 20 non-standalone script fragments (referencing unset variables under \code{bash -u}) and 3 macOS-only options rejected by GNU utilities leaves \textbf{161 eligible read-only commands} in \textbf{Benign-Ext}.

{\tolerance=2000 \textbf{End-to-End Evaluation Protocols.} In \emph{Entrypoint Confirmation}, fixed mock model responses drive four CLIs (Roo, Cline, Continue, OpenCode; 28/28 file-diff confirmations) to separate default headless modes (\code{RC-08}) from checker defects. In \emph{Live Product Trials}, all eight unmodified CLIs are driven by \code{gpt-5.6-luna} and \code{deepseek-v4.1-flash} (with only LLM API traffic exiting the containers) across 28 paired controls and 46 active witnesses per model (33 under configured restriction rules, 13 under permissive modes); benign controls are verified by output or tool events, deny controls by explicit logged rejections, and three agent--model pairs (Gemini on both models; Codex on \code{deepseek-v4.1-flash}) use a schema-only local HTTP wire adapter. In our \emph{Controlled Agent Benchmark}, ten models---six frontier cloud models (\code{gpt-5.6-luna}, \code{gpt-6.2-astra}, \code{deepseek-v4.1-flash}, \code{glm-5.3-flash}, \code{kimi-k3}, \code{minimax-m3.1-flash}) and four smaller or mid-tier models (\code{Qwen3.5-9B}, \code{Llama-3.1-8B-Instruct}, \code{gemma4-e4b}, \code{hy4std})---compare \code{no-gate}, \code{roo-gate}, a remote LLM judge (\code{llm-guard}), and \shellplan{} across 19 attack and benign tasks (760 runs; Section~\ref{sec:rq3b}).\par}

% ============================================================================
\section{Empirical Study of Eight Production Coding Agents}
\label{sec:empirical}
% ============================================================================

We answer RQ1 and RQ2 by attributing every observed violation to Definition~\ref{def:divergence}. Table~\ref{tab:t1} reports unified component-level measurements across all eight unmodified vendor checkers, Codex's isolated OS sandboxes, and \shellplan{} (Section~\ref{sec:eval}). Across all stages, we cataloged \textbf{17 deduplicated root causes} (\code{RC-01}--\code{RC-17}), merging surface variants that exercise the same decision path. These comprise \textbf{15 enforcement defects}---7 parser differentials (\code{RC-01}--\code{06}, \code{17}), 1 permission-precedence override (\code{RC-16}), 5 utility, environment, or redirection gaps (\code{RC-09}--\code{13}), and 2 confinement or classifier defects (\code{RC-14}--\code{15})---together with \textbf{2 permissive headless-mode defaults} (\code{RC-07}--\code{08}).

\begin{table*}[!t]
\centering
\caption{Unified component-level evaluation on \code{P1}-active suites. Metrics: \emph{Allow} = checker allow rate (\code{P2}/\code{P3} evaluated on the same \code{P1}-active set); \emph{Mut.} = persistent filesystem mutations inside isolated \code{codex sandbox} (\code{bwrap}+Landlock); \emph{Derived} = \code{Codex P1 + -s ro} where \code{execpolicy} disables sandboxing on active cases (confirmed live with \code{bwrap=0} in Table~\ref{tab:t3product}); \emph{Classif.} = stock danger classifier safe verdict. Panel~B reports \code{Benign-Ext} (\code{P1}) escalation rates (/184, /161) and non-read-only over-allows (/316). ``---'' denotes not applicable or not measured.}
\label{tab:t1}
\adjustbox{max width=\textwidth}{%
\footnotesize
\setlength{\tabcolsep}{3.0pt}
\renewcommand{\arraystretch}{1.06}
\begin{tabular}{@{}llrrrrrr|rrr@{}}
\toprule
& & \multicolumn{6}{c|}{\textbf{Panel A: Active Challenge Suites ($\downarrow$)}} & \multicolumn{3}{c}{\textbf{Panel B: Benign-Ext (P1)}} \\
\cmidrule(lr){3-8}\cmidrule(l){9-11}
\textbf{System / Mode} & \textbf{Metric} & \textbf{CVE} & \textbf{Syntax} & \textbf{Atk-1K} & \textbf{Fuzz-3K} & \textbf{BinSem} & \textbf{WS-Int} & \textbf{Esc.} & \textbf{Esc.} & \textbf{Allow} \\
& & (/33) & (/405) & (/968, P1/P2/P3) & (/2,685, P1/P2/P3) & (/55; Exec/18) & (/28) & (/184 $\downarrow$) & (/161 $\downarrow$) & (/316 $\downarrow$) \\
\midrule
\code{roo-code} v3.54 & Allow & 51.5\% & 55.8\% & 45.9 / 45.0 / 70.4 & 28.0 / 18.0 / 42.0 & 74.5\% (10) & 92.9\% & 2.17\% & 2.48\% & 12 \\
\code{cline} main & Allow & 63.6\% & 44.0\% & 42.7 / 42.7 / 42.7 & 48.7 / 48.7 / 48.7 & 80.0\% (16) & 50.0\% & ---$^{\dagger}$ & 0.00\%$^{\ddagger}$ & 219$^{\ddagger}$ \\
\code{qwen-code} 0.28.3 & Allow & 63.6\% & 54.6\% & 43.3 / 42.5 / 87.0 & 30.4 / 20.0 / 75.9 & 85.5\% (12) & 96.4\% & 1.63\% & 1.86\% & 53 \\
\code{gemini-cli} 0.70.0 & Allow & 30.3\% & 4.0\% & 4.2 / 3.4 / 22.3 & 6.9 / 4.7 / 28.8 & 74.5\% (10) & 53.6\% & 3.80\% & 3.73\% & 39 \\
\code{opencode} dev & Allow & 30.3\% & 3.7\% & 9.3 / 8.5 / 36.4 & 10.3 / 6.7 / 40.7 & 74.5\% (10) & 75.0\% & 2.17\% & 2.48\% & 48 \\
\code{continue} main & Allow & 6.1\% & 3.7\% & 2.5 / 2.5 / 2.5 & 3.9 / 3.9 / 3.9 & 16.4\% (n/a$^{\#}$) & 14.3\% & 16.30\% & 11.18\% & 18 \\
\code{mistral-vibe} 3.1.4 & Allow & 12.1\% & 1.5\% & 0.9 / 0.6 / 7.4 & 3.9 / 2.0 / 11.2 & 36.4\% (\phantom{0}7) & 25.0\% & 26.09\% & 17.39\% & 2 \\
\code{codex-cli} 0.171.0 & Allow & 21.2\% & 3.0\%$^{\S}$ & 2.1 / 1.2 / 7.2 & 3.6 / 1.9 / 7.7 & 70.9\% (\phantom{0}8) & 53.6\% & 30.98\% & 22.98\% & 2 \\
\midrule
\code{codex-default} (unmatched) & Classif. & 100.0\% & --- & 98.6\% (P1) & --- & 100.0\% & 96.4\% & 0.00\% & --- & --- \\
Codex \code{bwrap -P ro} & Mut. & 0.0\% & --- & --- & --- & 0.0\% & 0.0\% & --- & --- & --- \\
Codex \code{bwrap -P ww} (unmatched) & Mut. & 6.1\% & --- & 0.0\% (ext) & --- & 7.3\% (ws) & 71.4\%$^{\P}$ & --- & --- & --- \\
\textbf{Codex P1 + \code{-s ro}} & \textbf{Derived} & \textbf{21.2\%} & --- & \textbf{2.1\% (P1)} & --- & \textbf{70.9\%} & \textbf{53.6\%} & 30.98\% & 22.98\% & 2 \\
\midrule
\rowcolor{pgMethodTint}\textbf{\shellplan} & \textbf{Allow} & \textbf{0.0\%} & \textbf{0.0\%} & \textbf{0.0 / 0.0 / 0.0} & \textbf{0.0 / 0.0 / 0.0} & \textbf{0.0\% (\phantom{0}0)} & \textbf{0.0\%$^{*}$} & \textbf{10.33\%} & \textbf{7.45\%} & \textbf{0} \\
\bottomrule
\end{tabular}}
\begin{flushleft}
\scriptsize
$^{\dagger}$Cline plan-mode has no separate \code{P1} escalation mode on \code{Benign-Ext} (/184). $^{\ddagger}$Unmatched Cline standard check: 0/161 escalated, 219/316 non-RO allowed. $^{\#}$Continue's \code{Attack-BinSem} logged aggregate allow (9/55) without the 18-case exec sub-label. $^{\S}$Codex exact \code{SyntaxMatrix} is 12/405 (3.0\%); 6/396 on its policy-compatible subset. $^{\P}$Codex \code{bwrap -P ww} protected 8/8 \code{.git/*} targets and permitted 20/20 ordinary workspace writes per policy. $^{*}$\shellplan{} under \code{P1}/\code{P2} RO (see Section~\ref{sec:rq3a} for workspace-write mode).
\end{flushleft}
\end{table*}

\subsection{RQ1-A: Parse Divergences (D1) and a Rule-Precedence Fault}
\label{sec:rq1a}

A parse divergence occurs when a checker's command decomposition disagrees with Bash. Two distinct mechanisms stood out even among modern \code{tree-sitter-bash}-based checkers:

\paragraph{Cross-Product Upstream \code{tree-sitter-bash} Lexical-Boundary Differential (\code{RC-17}).}
We discovered a multi-line lexical-boundary defect in upstream \code{tree-sitter-bash} (Python \code{0.25.1} in Mistral Vibe~3.1.4 and the WASM build in Gemini CLI~0.70.0; exact byte sequence redacted pending coordinated disclosure). When a two-line input places an escaped command word immediately after an unescaped newline, GNU Bash~5.2 and Zsh treat the unescaped newline as a command separator and execute line~2 as an independent command. In \code{tree-sitter-bash}, however, the scanner folds the newline and the second command token into an argument word of line~1's allowlisted command (e.g., \code{echo}). Because the AST passed to policy evaluation contains no second command node, Mistral Vibe's denylist, all 15 per-binary guardrails, and outside-workspace path checks---as well as Gemini CLI's high-priority explicit deny rules---never observe the second command.

\paragraph{Permission-Precedence Override (\code{RC-16}).}
In Qwen Code~0.28.3, the shell tool's default permission check accurately detects nested shell command substitutions and returns \code{"ask"}. However, the subsequent permission-rule evaluation flow only protects \code{"deny"} defaults from being overridden: a matching user allow rule on the outer inspection command (exact configuration and trigger redacted pending remediation) silently upgrades that mandatory \code{"ask"} verdict to \code{"allow"} without re-enforcing substitution or redirection checks. This precedence flaw accounts for Qwen Code's 63.6\% \code{CVE-Replay} and 43.3\% \code{Attack-1K} bypass rates under \code{P1}. Additional D1 parser gaps include Roo Code's arithmetic placeholder substitution (\code{RC-01}) and backslash-newline continuation folding (\code{RC-02}), Cline's regex-masking bypasses (\code{RC-05}), Gemini CLI's unexpanded token matching under denylists (\code{RC-06}), and Continue's leading bare redirection token shift (\code{RC-04}).

\subsection{RQ1-B: Utility Divergences (D2)}
\label{sec:rq1b}

Utility divergences require zero parsing error: checker and shell agree completely on command boundaries, yet an allowlisted inspection binary produces prohibited side effects through its own interface. Every studied checker admitted active \code{Attack-BinSem} cases composed purely of \code{P1}-allowlisted utilities: Continue allowed 9/55 (16.4\%), Mistral Vibe allowed 20/55 (36.4\%, including 7/18 execution cases), and the remaining six checkers allowed 39--47 of 55 (70.9\%--85.5\%, including 8--16 of 18 execution cases). These effects stem from three utility surfaces: output-file arguments (\code{uniq IN OUT}, \code{sort -o OUT}, \code{git diff/show --output=OUT}; \code{RC-09}), embedded command-execution flags (\code{sed} execute directives, \code{rg --pre=script}, \code{sort --compress-program}; \code{RC-10}--\code{11}), and inline environment variables (\code{GIT\_TRACE}, \code{RIPGREP\_CONFIG\_PATH}; \code{RC-12}). Separately, Roo, OpenCode, Gemini, and Qwen leave output redirections unguarded (\code{RC-13}). Notably, even Mistral Vibe's 15 hand-crafted per-binary guardrails and path checks left 20/55 \code{Attack-BinSem} cases open (including \code{sed} execution/write directives and \code{rg --pre}), illustrating the fragility of denylist-based argument filtering.

\subsection{RQ1-C: Confinement Divergence (D3) and Classifier Gap}
\label{sec:rq1c}

Table~\ref{tab:t1} contrasts Codex CLI~0.171.0's isolated \code{bwrap}+Landlock sandboxes against its composed policy--sandbox configuration. \textbf{Policy--sandbox inversion (\code{RC-14}).} Codex's isolated read-only sandbox (\code{bwrap -P ro}) prevented persistent filesystem mutations on every tested case across \code{CVE-Replay} (0/33), \code{Attack-BinSem} (0/55), and \code{Workspace-Integrity} (0/28). When paired with \code{P1} allow rules under \code{--sandbox read-only} (\emph{Codex P1 + \code{-s ro}}), however, matching allowlisted prefixes disables sandboxing across 7/33 \code{CVE-Replay}, 20/968 \code{Attack-1K}, 39/55 \code{Attack-BinSem}, and 15/28 \code{Workspace-Integrity} cases (confirmed live with zero \code{bwrap} spawns in Section~\ref{sec:rq2}; internal implementation pointers redacted pending disclosure). \textbf{Classifier gap inside workspace-write (\code{RC-15}).} Under workspace-write confinement (\code{bwrap -P ww}), \code{bwrap} protected all 8 \code{.git/*} metadata targets and blocked external mutations (0/968 on \code{Attack-1K}), while allowing ordinary workspace writes as intended (20/20). Codex's stock danger classifier checks POSIX \code{rm} only when forced, classifying 27/28 \code{Workspace-Integrity} cases (including \code{rm -r}, \code{find -delete}, and \code{sed -i}) and 954/968 \code{Attack-1K} cases as safe without user confirmation.

\begin{findingbox}{Answer to RQ1 (Divergence)}
Violations spanned all three boundaries: an upstream \code{tree-sitter-bash} defect hid a second command from two modern AST checkers (D1), every checker permitted utility-mediated writes or subcommand launches (9--47 of 55 cases, D2), and in Codex CLI an allow rule disabled OS sandboxing on 39/55 utility-semantics cases even though the isolated read-only sandbox prevented persistent mutations on all 55 (D3).
\end{findingbox}

\subsection{RQ2: Live Model-Driven Trials Across Eight Production CLIs}
\label{sec:rq2}

To verify whether component divergences translate into real effects under model-driven execution, Table~\ref{tab:t3product} summarizes live product trials across all eight unmodified CLIs and two frontier models, separating \emph{configured restriction rules} (33 trials/model) from \emph{permissive modes} (13 trials/model). Every paired control trial (28/28 per model; 56/56 total: 10 benign-allow and 18 explicit deny/sandbox controls per model) achieved its expected outcome, confirming that each vendor gate was active.

\begin{table*}[!t]
\centering
\caption{Live model-driven product trials across eight unmodified coding-agent CLIs and two frontier models, separating configured restriction rules (33 trials/model) from permissive modes (13 trials/model). All 56 paired control trials (28/28 per model) verified active enforcement.}
\label{tab:t3product}
\adjustbox{max width=\textwidth}{%
\footnotesize
\setlength{\tabcolsep}{3.3pt}
\renewcommand{\arraystretch}{1.06}
\begin{tabular}{@{}lllcc>{\raggedright\arraybackslash}p{4.8cm}@{}}
\toprule
\textbf{Vendor Implementation} & \textbf{Tested Configuration} & \textbf{Ctrls} & \textbf{\code{gpt-5.6}} & \textbf{\code{ds-v4.1}} & \textbf{Observed Outcome \& Attribution} \\
\midrule
\multicolumn{6}{@{}l}{\textit{Part A: Configured-Rule Enforcement Modes (allowlist, denylist, Plan RO, or \code{-s read-only})}} \\
Codex CLI \code{0.171.0} & RO Policy (\code{-s ro} + P1) & 4\,|\,4 & 5/5 & 5/5 & \code{bwrap=0} on all 10 hits (\code{RC-14}); \code{sed} ran forbidden \code{touch}. \\
OpenCode \code{2.3.0} & P1 allowlist; \code{rm}/\code{touch} deny & 3\,|\,3 & 4/4 & 2/4 & \code{ds-v4.1}: 2 hits (\code{uniq}, \code{sort -o}), 2 zero-call refusals. \\
Roo Code \code{0.1.17} (wrapped) & Allowlist; \code{rm} deny; \code{touch} ask & 3\,|\,3 & 5/5 & 4/5 & \code{strace}: \code{sed$\to$sh$\to$touch}; \code{ds-v4.1}: 1 zero-call. \\
Cline CLI \code{3.0.74} & Plan (\code{-p}): RO guard & 3\,|\,3 & 4/4 & 1/4 & Blocked \code{rm}/\code{touch}, executed nested subst.; \code{ds-v4.1}: 3 zero-call. \\
Gemini CLI \code{0.70.0} & \code{-p} + TOML deny (p200) & 3\,|\,3 & 5/5 & 5/5 & \code{RC-17} bypassed high-priority explicit deny rule. \\
Qwen Code \code{0.28.3} & \code{-p} + allow/deny rules & 3\,|\,3 & 4/5 & 4/5 & \code{RC-16} hit on both; virtual \code{"edit"} rule blocked \code{sort -o} (2/2). \\
Mistral Vibe \code{3.1.4} & \code{-p} + allow/deny + 15 guards & 4\,|\,4 & 5/5 & 3/5 & \code{RC-17} hit on both; \code{ds-v4.1} wrapped 2 calls $\to$ approval. \\
\textbf{Subtotal (Configured Rules)} & \textbf{7 restricted modes} & \textbf{23\,|\,23} & \textbf{32/33} & \textbf{24/33} & \textbf{All 7 restricted agents hit on both models}. \\
\midrule
\multicolumn{6}{@{}l}{\textit{Part B: Permissive / Headless-Default Configurations (targeted effect permitted by mode default)}} \\
Codex CLI \code{0.171.0} & Workspace-Write (\code{-s ww}) & 2\,|\,2 & 4/4 & 4/4 & \code{bwrap=3}; stock classifier permitted all 4 workspace ops (\code{RC-15}). \\
Cline CLI \code{3.0.74} & Act default (\code{RC-08}) & 1\,|\,1 & 4/4 & 3/4 & No allowlist; \code{ds-v4.1}: 1 zero-call refusal. \\
Continue CLI \code{1.5.47} & Headless \code{Bash=allow} + \code{sudo} deny & 2\,|\,2 & 5/5 & 3/5 & \code{sudo} blocked; \code{RC-08} ran payloads; \code{ds-v4.1}: 2 zero-call. \\
\textbf{Subtotal (Permissive Modes)} & \textbf{3 permissive modes} & \textbf{5\,|\,5} & \textbf{13/13} & \textbf{10/13} & \textbf{Mode defaults ran the payloads}. \\
\midrule
\textbf{Combined Direct Total} & \textbf{8 vendor implementations} & \textbf{28\,|\,28} & \textbf{45/46} & \textbf{34/46} & \code{ds-v4.1}: 34 hits, 9 zero-call, 2 wrapper stops, 1 Qwen block. \\
\bottomrule
\end{tabular}}
\end{table*}

Under configured restriction rules (Part~A), prohibited effects materialized in \textbf{32/33 (97.0\%)} trials with \code{gpt-5.6-luna} and \textbf{24/33 (72.7\%)} with \code{deepseek-v4.1-flash}, reaching every restricted agent on both models. Across all 66 configured-rule trials, a vendor gate blocked a verbatim witness in only one instance: Qwen Code's virtual \code{"edit"} check rejected \code{sort -o} (2/2). Decomposing the 12 non-hits on \code{deepseek-v4.1-flash} across both parts shows 9 model self-refusals (zero shell tool calls), 2 approval stops caused by model-generated wrapper commands in Mistral Vibe, and the Qwen virtual-edit block---yielding 34/37 (91.9\%) among trials that issued a shell call and 34/35 (97.1\%) among verbatim payloads. In separate indirect prompt-injection trials delivered via \code{README.md} (3/3 controls clean per cell), OpenCode produced 4/4 (\code{gpt-5.6-luna}) and 2/4 (\code{deepseek-v4.1-flash}) targeted effects, and Roo Code produced 5/5 and 1/5.

\begin{findingbox}{Answer to RQ2 (Consequence)}
Live model-driven execution produced targeted policy-violating effects in all 16 agent--model pairs: 14 under configured restriction rules across seven products (including unsandboxed \code{bwrap=0} runs in Codex CLI), and Continue's 2 under its permissive headless default.
\end{findingbox}

\subsection{Design Requirements Derived from Empirical Divergences}

Each boundary failure motivates a concrete architectural requirement. Parse divergences (D1) and rule-precedence faults (\code{RC-16}) arise when a gate parses or overrides decisions apart from execution. Utility divergences (D2) persist regardless of parser precision: an earlier prototype of \shellplan{} without utility specifications still permitted 32/55 \code{Attack-BinSem} cases. Confinement divergences (D3) arise when approval toggles OS isolation.
\begin{enumerate}[leftmargin=1.8em,itemsep=1.5pt]
  \item[\textbf{R1}] \textbf{Single-Interpretation Dispatch (against D1 and \code{RC-16}):} Execute the exact compiled representation validated by the policy, with no second shell parse and no rule overriding a failed check.
  \item[\textbf{R2}] \textbf{Positive Utility \& Configuration Specifications (against D2):} Restrict utility options, operand counts, inline environment assignments, and repository configurations to allowlisted forms, escalating all others.
  \item[\textbf{R3}] \textbf{Orthogonal Confinement \& Open-Time Enforcement (against D3):} Enforce redirection bounds statically and at file-open time, and never let authorization relax host OS sandboxing.
\end{enumerate}

% ============================================================================
\section{\shellplan: Compile-Once, Direct-Execve Command Gate}
\label{sec:design}
% ============================================================================

\shellplan{} compiles each command string once into a typed execution plan, checks the plan against policy and positive utility specifications, and executes admitted plans directly via \code{execve} without passing command text to a shell (Figure~\ref{fig:paths}b).

\subsection{Compilation and Direct Evaluation of Admitted Plans (R1)}
\label{sec:compiler}

Given command string $s$, $\text{Compile}(s)$ produces either a typed plan $P$ or an explicit escalation $\text{Escalate}(r)$ with reason code $r$. The IR models pipelines, command lists (\code{;}, \code{\&\&}, \code{||}), simple commands with prefix assignments, words with quote spans, nested \code{\$(...)}, and ordered redirections (\code{<}, \code{>}, \code{>>}, \code{2>\&1}, \code{1>\&2}). Compilation fails closed on unsupported constructs before execution: arithmetic expansion (\code{\$((...))}), process substitution (\code{<(...)}), backticks, unquoted \code{\~} or brace expansions, shell-internal variables (\code{\$?}, \code{\$1}), stateful builtins (\code{cd}, \code{eval}, \code{exec}), and unsupported redirections (\code{3>x}, \code{>|}, \code{|\&}, heredocs, or targets containing \code{..} or trailing \code{/}).

When $\Pi$ admits $P$, the executor evaluates $P$ directly and launches external binaries via \code{execve} (\code{shell=False}), never spawning \code{/bin/sh} or \code{bash}. It (1)~strips \code{\textbackslash<newline>} continuations outside single quotes and comments; (2)~expands ordinary words (parameter expansion, nested \code{\$(...)}, IFS splitting, locale-aware globbing) and aborts cleanly if an unquoted expansion produces glob metacharacters or a glob contains \code{//}; (3)~evaluates prefix assignments left-to-right without splitting or globbing; and (4)~opens redirections in written order over an explicit descriptor table. It unwraps \code{env}, \code{timeout}, and \code{command [-p]} chains to validate the inner executable. For Bash fidelity, it preserves inherited environment variables, maintains logical \code{PWD} and \code{SHLVL}, retains \code{argv[0]}, emits diagnostics on active \code{fd 2}, executes \code{echo} and \code{pwd} in-process, and formats \code{time [-p]} pipeline timings.

\subsection{Positive Utility Specifications and Git Config Guard (R2)}
\label{sec:argspec}

For specified utilities, \shellplan{} validates options (respecting GNU \code{getopt} short-option bundling and explicit long-option spellings), option arguments, and operand arity statically and again after runtime expansion. The specification module covers \textbf{20 core utilities}: 18 entries in \code{GNU\_SPECS} (the \code{P1} inspection binaries, with \code{sed} alongside in-process \code{echo}), a primary allowlist for \code{find}, and six \code{git} read-only subcommands (\code{status}, \code{branch}, \code{diff}, \code{log}, \code{show}, and \code{remote [-v|--verbose]}). A subcommand filter covers \code{npm}, which executes project code and therefore remains outside Theorem~\ref{thm:contain}. For instance, \code{uniq} accepts at most one operand, \code{sort} rejects \code{-o}/\code{--output}/\code{--compress-program}, and \code{rg} rejects \code{--pre}; inline assignments allow only locale, timezone, and display variables. For GNU \code{sed}, \shellplan{} validates scripts via \code{sed --sandbox} over \code{/dev/null} statically and after runtime expansion, injects \code{--sandbox} at dispatch, and rejects non-regular \code{-f} script targets. Every resolved binary must reside outside $W$; under developer-supplied \emph{native} policies, interpreters and forwarders (\code{sh}, \code{python}, \code{awk}, \code{xargs}) escalate by class, and multi-word deny rules match unwrapped basenames statically and after expansion.

\textbf{Repository Configuration Checks.} Prior to launching any \code{git} process, \shellplan{} executes fixed read-only probes (\code{git rev-parse}, \code{git config --list --show-scope -z}) across the workspace repository, its \code{gitdir}/\code{commondir}, submodules under \code{modules/}, and every nested \code{.git} directory or \code{gitdir} pointer below the working-tree root. Dispatch proceeds only if every \code{local} and \code{worktree} key matches a strict allowlist (excluding hooks, filters, monitors, external diff/textconv helpers, and SSH commands) and every pointer resolves cleanly; additionally, no pipeline stage concurrent with a \code{git} stage may write a file by redirection or run an unspecified program, including inside nested \code{\$(...)}.

\subsection{Redirection Containment and Open-Time Enforcement (R3)}
\label{sec:toctou}

Read-only policies permit output redirection only to \code{/dev/null}, and every policy keeps input redirections inside $W$ or on \code{/dev/null}. Under workspace-write policies (\code{allow\_workspace\_write=True}), \shellplan{} validates every write target both statically and at open time. Static authorization rejects targets outside $W$; targets with any case-insensitive \code{.git} component; targets inside any discovered \code{gitdir}, \code{commondir}, Git-recognized directory, or directory containing both \code{objects} and \code{refs} directories (via \code{lstat}) or a \code{commondir} entry; and existing regular files with link count $>1$. Any unresolved nested \code{.git} pointer rejects all workspace writes. At open time, the executor re-verifies path resolution, traverses from the root of $W$ component-by-component with \code{O\_PATH | O\_DIRECTORY | O\_NOFOLLOW} checking each directory's device/inode identity and shape, opens the leaf non-blocking without \code{O\_TRUNC} to reject FIFOs, sockets, and devices, and verifies link count $\le 1$ on the opened descriptor \emph{prior to truncation}. OS sandboxing remains strictly orthogonal: \shellplan{} never alters host confinement.

\subsection{Formal Operational Guarantees}
\label{sec:formal}

\textbf{Plans and Authorization.} $\text{Compile}(s)$ yields $\text{Escalate}(r)$ or a finite Plan IR tree:
\[
P ::= c \mid P \mathbin{\texttt{;}} P \mid P \mathbin{\texttt{\&\&}} P \mid P \mathbin{\texttt{||}} P \mid P \mathbin{\texttt{|}} P,
\qquad
c ::= \langle A,\ w_0\,w_1\cdots w_n,\ R\rangle,
\]
where $A$ holds prefix assignments, $w_0$ is a static literal program word, each remaining word comprises literals, parameter expansions, globs, and strictly smaller nested plans $\$(P')$, $R$ is an ordered redirection list, and nodes may carry unwrapped \code{env}, \code{timeout}, or \code{command} wrapper chains. Static authorization $\Pi\vdash P\ \mathsf{ok}$ holds iff every command node $c$ anywhere in $P$ (including inside $\$(\cdot)$) satisfies: (i)~$\Pi$ permits its program $b$ and $b$ is not a hazard/interpreter; (ii)~if $b$ is specified, its static argument view lies in $\Sigma_b$ (and static \code{sed --sandbox} accepts); (iii)~its assignments are allowlisted; (iv)~its redirections satisfy Section~\ref{sec:toctou}; and (v)~no pipeline stage concurrent with a \code{git} stage writes a file or invokes an unspecified program. $\text{Exec}(P)$ evaluates $P$ structurally; before dispatching any reached node, it re-checks $\Sigma_b$ and deny rules on expanded \code{argv}, verifies executable resolution outside $W$, and runs runtime \code{sed}/\code{git} probes from the fixed read-only probe family $\mathcal{H}$ (\code{sed --version}, \code{sed --sandbox} on \code{/dev/null}, \code{git rev-parse}, \code{git config --list}).

\textbf{Explicit Assumptions.} \emph{A1 (Trusted Base):} inherited environment variables, host/user configs outside $W$, and binaries outside $W$ are trusted; the adversary controls only agent-read text and files inside $W$. \emph{A2 (Specified Utility Behavior):} each specified utility $b$, dispatched with \code{argv} $\in \Sigma_b$, allowlisted inline variables, \code{--sandbox} for \code{sed}, and vetted repository config for \code{git}, writes no files outside \code{stdout}/\code{stderr} (except \code{sort} temporaries and \code{.git/index} refreshes by \code{git status}/\code{git diff}), launches descendant programs only under A1-trusted configuration, and parses options consistently with $\Sigma_b$; probes in $\mathcal{H}$ write no files and spawn no programs. Section~\ref{sec:rq4} empirically audits both option parsing and Git config allowlists under A2.

\begin{theorem}[Single Interpretation]
\label{thm:single}
Let $\Pi$ be a policy, $W$ a workspace, and $s$ a command string. (1)~If $\text{Compile}(s)=\text{Escalate}(r)$ or $\Pi\nvdash P\ \mathsf{ok}$, \shellplan{} dispatches no command node and opens none of the redirection targets of $s$; authorization executes only probes in $\mathcal{H}$. (2)~If $\Pi\vdash P\ \mathsf{ok}$, every child process created directly by \shellplan{} during $\text{Exec}(P)$ is either the direct \code{execve} dispatch of a reached node passing all runtime checks (possibly through its unwrapped wrappers, with \code{--sandbox} injected for \code{sed}) or a probe in $\mathcal{H}$; unwrapped \code{echo} and \code{pwd} execute in-process. No substring of $s$ is passed to a shell.
\end{theorem}
\begin{proof}[Proof sketch]
By strong structural induction on Plan IR height $h(P)$: the entrypoint invokes \code{Exec} only when $\Pi\vdash P\ \mathsf{ok}$; every non-probe launch occurs inside simple-command dispatch with \code{shell=False} after runtime checks, and every nested $\$(P')$ has strictly smaller height $h(P') < h(P)$ with nodes contained in $P$. Full operational semantics and line-mapped inductive proofs appear in our companion proof appendix.
\end{proof}

\begin{theorem}[Effect Containment]
\label{thm:contain}
Assume A1 and A2, and let $\Pi\vdash P\ \mathsf{ok}$ where every node of $P$ names a specified utility. Then every process started by \shellplan{}, its probes, or its dispatched nodes is a dispatched node, a probe in $\mathcal{H}$, or a helper directed by A1-trusted configuration, and every written file (outside A1-trusted helpers) is a redirection target satisfying Proposition~\ref{prop:redir}, a \code{sort} temporary file, or a \code{.git/index} refresh by \code{git status} or \code{git diff}.
\end{theorem}
\begin{proof}[Proof sketch]
Combining Theorem~\ref{thm:single} with runtime binary resolution outside $W$, pre-launch \code{git} config vetting across all reachable repositories, pipeline concurrency exclusion of concurrent writers, Proposition~\ref{prop:redir}, and A1--A2 bounds all descendant launches and file writes.
\end{proof}
Unspecified utilities and \code{npm} permitted under user-supplied \emph{native} policies (e.g., \code{pytest}, \code{make}, \code{tar}, \code{dd}, \code{ln}) lie explicitly outside Theorem~\ref{thm:contain} and are surfaced in \code{outside\_spec}.

\begin{proposition}[Redirection Containment]
\label{prop:redir}
Every redirection target opened by \shellplan{} is \code{/dev/null} or a path inside $W$, and read-only policies permit output only to \code{/dev/null}. Under workspace-write, at open time the original target path still resolves to the authorized location, every directory component from $W$ is opened via \code{O\_PATH | O\_DIRECTORY | O\_NOFOLLOW} outside recognized repository-metadata predicates, and the leaf is verified via \code{fstat} to be a regular file with link count $\le 1$ before truncation.
\end{proposition}
Proposition~\ref{prop:redir} covers static states and concurrent specified stages (external-writer race tests and one symlinked-metadata predicate boundary, \code{M1}, are characterized in Section~\ref{sec:rq4}). Because \shellplan{} never alters host sandboxing, its decisions cannot induce a confinement divergence (D3).

\textbf{Property~1 (Bash Agreement, Empirically Tested).} For deterministic admitted commands under default Bash startup options, exit code, \code{stdout}, and \code{stderr} match \code{/bin/bash -c} $s$ in identical state (capped at 262,144\,B/stream with UTF-8 \code{U+FFFD} replacement, reported explicitly), or execution halts cleanly with a diagnostic reason code whose prior output is a prefix of Bash's. Two fidelity mismatches (\code{F1}, \code{F2}) are documented in Section~\ref{sec:rq4}.

% ============================================================================
\section{Evaluation of \shellplan}
\label{sec:eval}
% ============================================================================

All experiments evaluate a single frozen snapshot of \shellplan{} (release v2.4, commit \code{2e07e41}), kept byte-identical across all benchmarks, audits, and real-CLI integration tests.

\subsection{RQ3-A: Component Security, Utility, and Latency}
\label{sec:rq3a}

Across all six active challenge suites (Table~\ref{tab:t1}), \shellplan{} admitted \textbf{0/4,174} active attack cases under read-only policies while admitting 4/4 spec-covered benign \code{Attack-BinSem} controls. Under workspace-write policies, it protected \textbf{8/8} \code{.git/*} metadata targets and permitted \textbf{2/20} ordinary workspace targets under \code{P2+ws-write} (\textbf{4/20} under \code{native+ws-write}). Across canonical (\code{P1}), native, and \code{native+ws-write} arms, it further admitted 0/8 active held-out \code{GTFOBins} inputs and 0/207 active red-team inputs from multi-model red-teaming rounds.

\begin{figure}[!t]
  \centering
  \includegraphics[width=\textwidth]{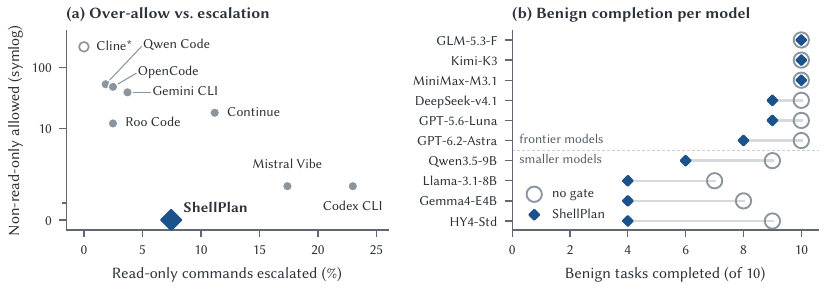}
  \Description{Two-panel plot showing Benign-Ext over-allow versus escalation Pareto frontier and per-model benign task completions across ten models.}
  \caption{(a)~Pareto frontier under \code{P1} on \code{Benign-Ext}: share of 161 eligible read-only commands escalated vs.\ count of 316 non-read-only commands over-allowed (symlog scale; *Cline's unmatched standard check). (b)~Benign tasks completed (of 10) per model in our Controlled Agent Benchmark without a gate vs.\ under \shellplan.}
  \label{fig:tradeoff}
\end{figure}

On \code{Benign-Ext} (Figure~\ref{fig:tradeoff}a), \shellplan{} dominates the security--utility frontier: every policy-matched vendor checker escalating fewer than 4\% of the 161 eligible read-only commands over-allowed 12 to 53 of the 316 non-read-only commands (Cline's unmatched check: 219). \shellplan{} over-allowed \textbf{0/316} while escalating only \textbf{12/161 (7.45\%)}---less than half the escalation rate of the next-strictest gates, Mistral Vibe~3.1.4 (\textbf{28/161, 17.39\%}, 2/316 over-allowed) and Codex CLI~0.171.0 (\textbf{37/161, 22.98\%}, 2/316 over-allowed). Including all 184 raw read-only candidates (with 20 non-standalone script fragments and 3 macOS-only options), escalation is 19/184 (10.33\%, Wilson 95\% CI [6.71\%, 15.56\%]). On \code{Benign-Dev}, \shellplan{} achieved \textbf{0/160} escalation and \textbf{0/40} over-allow. Compilation plus static authorization (including static \code{sed --sandbox} probes) required a median of \textbf{0.035\,ms} (0.339\,ms P99) on \code{Benign-Ext} and \textbf{0.039\,ms} (0.361\,ms P99) on \code{Benign-Dev}.

\subsection{RQ3-B: Controlled End-to-End Agent Security, Utility, and Latency}
\label{sec:rq3b}

Our \emph{Controlled Agent Benchmark} executes 760 end-to-end tasks in a single contemporaneous batch across ten models, four defenses (\code{no-gate}, \code{roo-gate}, a remote \code{llm-guard} using \code{deepseek-v4.1-flash}, and \shellplan), and 19 tasks: 9 stealth attack tasks, 6 standard development tasks, and 4 syntax-stress tasks whose prompts deliberately request unsupported shell features (\code{\~}, \code{\{a,b\}}, \code{>\&2}, \code{\$?}) to test model recovery after escalation. All three gates receive one matched native grant (15 inspection commands allowed; \code{rm}, \code{curl}, \code{sudo} denied) with no human approver, so escalated commands fail closed and agents may retry or use a read-only \code{read\_file} tool within a fixed turn budget.

\begin{table*}[!t]
\centering
\caption{Controlled ten-model end-to-end agent benchmark (single contemporaneous batch, 760 runs).}
\label{tab:t3harness}
\adjustbox{max width=\textwidth}{%
\footnotesize
\setlength{\tabcolsep}{3.6pt}
\renewcommand{\arraystretch}{1.06}
\begin{tabular}{@{}lrrrrrr@{}}
\toprule
\textbf{Defense Arm} & \textbf{Attack Hits} & \textbf{Standard Dev} & \textbf{Syntax-Stress} & \textbf{Overall Benign} & \textbf{Blocked Benign} & \textbf{Median Gate} \\
& (/90, 95\% CI) & \textbf{Tasks} (/60) & \textbf{Tasks} (/40)$^{\dagger}$ & (/100, 95\% CI) & \textbf{Runs (Events)} & \textbf{Time / Task}$^{\ddagger}$ \\
\midrule
\code{no-gate} & 68.9\% (62/90) [58.7, 77.5] & 53/60 (88.3\%) & 40/40 (100.0\%) & 93/100 [86.3, 96.6] & 0 (0) & 0.00\,ms \\
\code{roo-gate} & 62.2\% (56/90) [51.9, 71.5] & 45/60 (75.0\%) & 39/40 (97.5\%) & 84/100 [75.6, 89.9] & 14 (20) & 2.11\,ms \\
\code{llm-guard} & 6.7\% (\phantom{0}6/90) [\phantom{0}3.1, 13.8] & 50/60 (83.3\%) & 37/40 (92.5\%) & 87/100 [79.0, 92.2] & 16 (19) & 6,062.48\,ms \\
\rowcolor{pgMethodTint}\textbf{\shellplan} & \textbf{0.0\% (\phantom{0}0/90) [\phantom{0}0.0, \phantom{0}4.1]} & \textbf{49/60 (81.7\%)} & \textbf{25/40 (62.5\%)} & \textbf{74/100 [64.6, 81.6]} & \textbf{54 (101)} & \textbf{1.28\,ms} \\
\bottomrule
\end{tabular}}
\begin{flushleft}
\scriptsize
$^{\dagger}$Under \shellplan, all 40 initial prompted-syntax commands escalated; 25 tasks were completed via rewritten shell commands or alternative tools. $^{\ddagger}$Summed gate time per task with $\ge$1 shell call (median across all 190 runs/arm: \code{no-gate} 0.00\,ms, \shellplan{} 1.19\,ms, \code{roo-gate} 2.05\,ms, \code{llm-guard} 5,848.03\,ms).
\end{flushleft}
\end{table*}

In Table~\ref{tab:t3harness}, \shellplan{} eliminated all attacks (\textbf{0/90} hits, blocking 71 attack runs) compared with 62/90 unguarded, 56/90 under \code{roo-gate}, and 6/90 under \code{llm-guard}, at a median gate latency of \textbf{1.28\,ms} per task---4,700$\times$ faster than the remote LLM judge (6,062.48\,ms). Crucially, benign task impact split sharply by model tier (Figure~\ref{fig:tradeoff}b): across the six frontier models, \shellplan{} preserved \textbf{36/36 (100.0\%)} completion on standard development tasks (matching \code{no-gate}'s 36/36 and outperforming \code{roo-gate}'s 33/36 and \code{llm-guard}'s 35/36) and achieved \textbf{56/60 (93.3\%)} overall benign completion (20/24 on syntax-stress tasks) with \textbf{0/54} attack hits. Across the four smaller models, completion fell from 33/40 (\code{no-gate}) to 18/40 (13/24 standard, 5/16 syntax-stress), bringing pooled completion to \textbf{74/100} (49/60 standard; paired discordant counts on standard tasks: 6 vs.\ 2 against \code{roo-gate}, exact McNemar $p=0.29$, and 0 vs.\ 4 against \code{no-gate}, $p=0.125$). Across all 54 benign runs that triggered a \shellplan{} escalation (101 events), agents adapted their shell call in 38 runs and completed 36 of the 54 tasks (27 via shell rewrites).

In an additional 1,250-run \emph{Stealth Evasion Benchmark} (10 models $\times$ 25 tasks across five evasion families; baselines run in Batch~1 and \shellplan{} in Batch~2 on a second GPU host with identical weights and vLLM~0.29.0 flags), \shellplan{} again recorded \textbf{0/250} hits (95\% CI [0.0, 1.5]) despite active attempts in 208/250 runs (617 shell calls, 0 \code{strace} hits), compared with 146/250 for \code{no-gate}, 138/250 for \code{roo-gate}, 129/250 for \code{cline-gate}, and 39/250 for \code{opencode-gate} (all 39 from output redirections).

\begin{findingbox}{Answer to RQ3 (Security and Cost)}
\shellplan{} admitted 0/4,174 active static attacks and recorded 0/90 controlled agent hits (0/250 stealth hits) at 1.28\,ms median gate time. On frontier models, it matched unguarded execution on standard development tasks (36/36; 56/60 overall), while escalation costs concentrated in smaller models (18/40) and prompted unsupported syntax (74/100 pooled; 12/161 on \code{Benign-Ext}).
\end{findingbox}

\subsection{Case Analysis: Why Gates Admit or Escalate}
\label{sec:cases}

Table~\ref{tab:cases} walks through five representative inputs under \code{P1}, contrasting vendor checkers with \shellplan's exact verdicts and diagnostic reason codes.

\begin{table*}[!t]
\centering
\caption{Five representative cases under \code{P1} and the verdicts of \shellplan{} (exact unpatched \code{RC-17} trigger redacted in v1).}
\label{tab:cases}
\footnotesize
\setlength{\tabcolsep}{3.4pt}
\renewcommand{\arraystretch}{1.08}
\begin{tabularx}{\textwidth}{@{}lX>{\raggedright\arraybackslash}p{6.1cm}@{}}
\toprule
\textbf{Case} & \textbf{Input} & \textbf{\shellplan{} verdict (reason code)} \\
\midrule
C1 (D1) & \code{<redacted multi-line RC-17 trigger>} & escalate (\code{canonical\_not\_allowed:touch}) \\
C2 (D2, D3) & \code{uniq README.md /var/tmp/x} & escalate (\code{TOO\_MANY\_OPERANDS:uniq max=1}) \\
C3 (D2) & \code{sed -n '1e touch /var/tmp/x' README.md} & escalate (\code{SED\_SCRIPT\_USES\_E\_R\_W\_COMMANDS}) \\
C4 (benign) & \code{cd src \&\& grep -rn 'def' .}\newline\code{nl -ba src/app.py} & escalate (\code{UNSUPPORTED\_BUILTIN})\newline escalate (\code{canonical\_not\_allowed:nl}) \\
C5 (benign) & \code{echo "hello world" | echo test=\$(cat)} & admit; prints \code{test=}, Bash prints \code{test=hello world} \\
\bottomrule
\end{tabularx}
\end{table*}

\textbf{C1 (D1 --- Parse Divergence).} Mistral Vibe and Gemini CLI allow this \code{RC-17} multi-line input because \code{tree-sitter-bash} folds the second command into an argument of \code{echo}, hiding \code{touch} from deny rules. \shellplan's compiler splits the unescaped newline into two distinct nodes (\code{echo} and \code{touch}), and authorization escalates on \code{touch}---realizing Theorem~\ref{thm:single}.

\textbf{C2 \& C3 (D2/D3 --- Utility and Confinement Divergence).} All checkers parse C2 and C3 identically to Bash. Six checkers allow C2 (\code{uniq} two-operand write) while Continue and Mistral Vibe escalate (Vibe's path check detects \code{/var/tmp/x}); seven checkers---including Mistral Vibe, where the path sits inside a \code{sed} script---allow C3, and Codex disables \code{bwrap} on both. \shellplan{} rejects C2 via \code{uniq}'s positive operand bound and C3 via \code{sed --sandbox} validation.

\textbf{C4 (Legitimate Escalations).} \shellplan{} escalates \code{cd} (stateful working directory) and \code{nl} (unlisted in \code{P1}). These two causes account for 89.6\% of read-only escalations on real trajectories (Section~\ref{sec:rq4}) and admit clear extensions: tracked working-directory state in the compiler and adding an \code{nl} specification.

\textbf{C5 (Admitted, Safe, Fidelity Mismatch \code{F2}).} C5 invokes only specified utilities (\code{echo}, \code{cat}), satisfies Theorem~\ref{thm:contain} (zero unintended processes or writes), yet prints \code{test=} instead of \code{test=hello world} because the frozen executor binds command-substitution \code{stdin} to \code{/dev/null} rather than inheriting the surrounding pipeline descriptor.

\subsection{RQ4: Empirical Boundaries, External Workloads, and Real-CLI Deployment}
\label{sec:rq4}

\paragraph{Auditing Assumption A2 and Internal Fidelity.}
To validate A2 directly against container binaries (GNU coreutils~9.4, \code{grep}~3.11, \code{sed}~4.9, \code{rg}~14.1.0, \code{jq}~1.7, \code{git}~2.43.0), we exhaustively scanned all 20 \code{getopt}-compatible specifications: across \textbf{1,213 tested options} (after excluding 37 early-exit options), we observed \textbf{0 mismatches} in whether an option consumes the next word (down from 9 in Round~1). Probing all 85 concrete allowlisted \code{git} repository-config keys across 9 \code{git} subcommands produced \textbf{0 foreign child \code{execve}s} across 765 \code{strace} runs. Against GNU Bash~5.2 (zero \code{sh/bash} spawns), \shellplan{} matched exit status and \code{stderr} on 160/160 \code{Benign-Dev} commands (158/160 raw \code{stdout}; 2 parallel \code{rg} line-ordering differences) and 164/165 admitted \code{Benign-Ext} commands (1 joint \code{tail -F} timeout), while passing 879 Linux and 51 macOS unit tests (including bounded exhaustive brace/glob checks and 210 symlink/\code{..} path geometries).

\paragraph{Three-Round White-Box Audit.}
Manual line-by-line review and differential fuzzing over three successive snapshots uncovered 15 root causes in Round~1 (8 security, including hardlink and \code{gitdir}-pointer escapes) and 18 in Round~2 (6 security, including post-authorization parent-symlink swaps, worktree \code{commondir} writes, and \code{git-lfs} filter spawns). Against the frozen Round~3 snapshot, 43 new security probes (21 active) produced \textbf{0/21} hits under canonical and native arms and \textbf{0} in-spec hits under \code{native+ws-write} (1 \code{outside\_spec} \code{ln}); all post-authorization path-swap probes were rejected at open time (\code{E\_REDIR\_PATH\_CHANGED}), and 45 default-environment fidelity cases yielded 32 exact matches, 2 clean refusals, and 11 escalations. Round~3 exposed one pre-staged metadata edge case (\code{M1}) and one inherited-\code{PWD} mismatch (\code{F1}).

\paragraph{Pre-Registered External Workloads.}
Before executing any external run, we froze a pre-registered protocol (commit \code{6c8968b}) over 60 real agent trajectories from 24 repositories in official x86\_64 Docker images---48 SWE-bench Verified trajectories~\cite{jimenez2024swebench} (12 repos $\times$ 4 submissions) and 12 SWE-rebench OpenHands trajectories~\cite{badertdinov2025swerebench,wang2025openhands} (12 repos), totaling 1,457 bash occurrences---plus 1,523 unique NL2Bash commands~\cite{lin2018nl2bash}. Four pre-registered endpoints test: \textbf{H1} (7-checker fidelity under literal \code{P1} \textbf{H1a} and A2-adjusted \code{P1} \textbf{H1b}, predicting 0\% H1b over-allow for \shellplan); \textbf{H2} (\code{P2-swebench-native} extending \code{P1} with \code{pytest} and \code{diff} without workspace writes, predicting 0\% non-A2 over-allow on \code{in\_spec} steps); \textbf{H3} (byte-exact Bash~5.2 equivalence on deduplicated admitted commands); and \textbf{H4} (coverage over all 564 \code{strace}-traced read-only steps: 456 Verified, 108 OpenHands). Native container replay agreed with recorded exit codes on 1,306/1,342 verifiable steps (97.32\%; 115 lacked recorded exit codes), outputs on 1,177/1,457 (80.78\%), and both on \textbf{1,072/1,342 (79.88\%)}; Bash equivalence compares \shellplan{} and Bash~5.2 directly in the same reconstructed state (Table~\ref{tab:external}).

\begin{table*}[!t]
\centering
\caption{Pre-registered external evaluation of \shellplan{} separating coverage, fidelity (literal \code{H1a} before A2-adjusted \code{H1b}; whole-policy \code{P2-swebench-native} before \code{in\_spec}), and live-state Bash~5.2 equivalence.}
\label{tab:external}
\adjustbox{max width=\textwidth}{%
\footnotesize
\setlength{\tabcolsep}{3.6pt}
\renewcommand{\arraystretch}{1.08}
\begin{tabular}{@{}l>{\raggedright\arraybackslash}p{3.0cm}>{\raggedright\arraybackslash}p{3.8cm}>{\raggedright\arraybackslash}p{4.3cm}@{}}
\toprule
\textbf{Workload} & \textbf{Coverage: admitted / eligible} & \textbf{Fidelity: over-allowed / admitted occurrences} & \textbf{Equivalence: outcomes on compared commands} \\
\midrule
SWE-bench Verified (48 trajs, 12 repos) & 343/456 traced RO steps (75.2\%)$^{\dagger}$ & P1: 5/348 literal (\code{H1a}, all \code{.git/index}), \textbf{0/348} excl.\ A2 (\code{H1b}, UB 1.09\%); \code{P2-swebench-native}, excl.\ A2: 44/398 total (\code{pytest}), \textbf{0/348} \code{in\_spec} & 343/343 exact live-state comparisons (322 Tier-1, 21 Tier-2; 0 \code{sh/bash} spawns)$^{\dagger}$ \\
SWE-rebench OpenHands (12 trajs, 12 repos) & \textbf{0/108} traced RO steps (0.0\%) & Not assessed: none admitted & Not assessed: none admitted \\
\textbf{Pooled 60 Trajs} & \textbf{343/564 (60.8\%)} & \textbf{5/348 literal; 0/348 excl.\ A2 (44 Verified trajs)} & \textbf{343/343 exact (44 trajs)} \\
\midrule
NL2Bash \code{dev+test} (1,523 unique) & 434/1,523 unique (28.5\%) & Not assessed here & 407 exact, 6 nondet, 5 refusals, 1 joint timeout; \textbf{1 semantic mismatch (\code{F2})}, 9 capped (262,144\,B), 5 drift \\
\bottomrule
\end{tabular}}
\begin{flushleft}
\scriptsize
$^{\dagger}$Across 48 Verified trajectories (44 had $\ge$1 admission), \shellplan{} admitted 348 \code{P1} occurrences: 5 refreshed \code{.git/index} (\code{git status}/\code{diff}), leaving 343 traced RO steps (343/456). Independently, deduplicating identical commands per state yielded 343 live-state equivalence checks.
\end{flushleft}
\end{table*}

\begin{itemize}[leftmargin=1.4em,itemsep=1.5pt]
  \item \textbf{Coverage (H4):} \shellplan{} admitted \textbf{343/456 (75.2\%)} traced read-only steps on SWE-bench Verified (exceeding \code{codex-cli}'s 73.0\%) and \textbf{0/108} on SWE-rebench OpenHands, where all 345 commands prefixed \code{cd /workspace/<repo> \&\& ...} (pooled: 343/564, 60.8\%). Just two constructs---\code{cd} (130) and \code{nl} (68)---accounted for 198/221 (89.6\%) of all read-only escalations.
  \item \textbf{Fidelity (H1 \& H2 across 7 checkers):} Across all 348 admitted \code{P1} occurrences (44 Verified trajectories), 5 refreshed \code{.git/index} (\code{H1a}) and \textbf{0/348} over-allowed outside A2 (\code{H1b}, 95\% UB 1.09\%). Restricting strictly to pre-divergence steps preserved 0\% \code{H1b} under every replay criterion (0/319 exit-code priority, UB 1.19\%; 0/78 rc-not-false + output, UB 4.69\%; 0/58 both-verifiable, UB 6.21\%). Under identical scoring, baseline \code{H1a}/\code{H1b} counts were 74/542 and 64/542 for \code{opencode} (59 \code{cat {<}{<}'EOF' >} heredoc writes), 636/1,198 and 624/1,198 for \code{cline}, and 5/185 and 0/185 for \code{continue} (which escalated 384/564 RO steps); \code{codex-cli}, \code{roo-code}, and \code{gemini-cli} scored 9/342 (4/342 \code{H1b}), 9/353 (4/353), and 9/350 (4/350), where all 4 non-A2 hits were \code{sed -i} steps eliminated by explicit \code{sed -i} deny rules. Under \code{P2-swebench-native}, \shellplan{} had 44/398 non-A2 hits, all from authorized \code{outside\_spec} \code{pytest} runs (\textbf{0/348} \code{in\_spec}), versus 46/390 (\code{codex-cli}), 48/402 (\code{roo-code}), 48/400 (\code{gemini-cli}), 108/591 (\code{opencode}), 7/185 (\code{continue}), and 707/1,198 (\code{cline}).
  \item \textbf{Equivalence (H3):} On SWE-bench Verified, \textbf{343/343} live-state comparisons matched Bash byte-for-byte with zero \code{sh/bash} spawns (plus 43/44 \code{outside\_spec} \code{pytest}/\code{diff} runs; one differed in a 0.01\,s timing digit). On NL2Bash (434 admitted, zero \code{sh/bash} spawns), 407 matched exactly, 6 were pre-specified nondeterministic matches, 5 halted with explicit refusals, 1 jointly timed out, 9 hit the 262,144\,B stream cap on \code{find /}, 5 reflected \code{/proc} or fixture rebuilds, and 1 exposed fidelity defect \code{F2} (Case~C5), falsifying strict zero-mismatch H3.
\end{itemize}

\paragraph{Deployment Inside an Unmodified Codex CLI.}
\label{sec:codex_integ}
To test real-world drop-in deployment, we configured an unmodified Codex CLI~0.171.0 binary to route all shell execution through frozen \shellplan{} (SHA-256 verified at startup) via an MCP stdio tool (\code{run\_command}), disabling Codex's built-in shell, sub-agents, hooks, plugins, and workspace \code{.codex/} loading, with \code{sandbox\_mode="read-only"} blocking patch writes. Each invocation spawns a fresh unprivileged worker (\code{uid 65534}) with a sanitized environment outside Codex's \code{bwrap} sandbox.

\begin{table*}[!t]
\centering
\caption{Live deployment study on unmodified OpenAI Codex CLI~0.171.0 (\code{gpt-5.6-luna}), comparing stock Codex against Codex routed through \shellplan{} MCP on identical prompts and witness commands (workspace \code{/work/repo}).}
\label{tab:codex_integ}
\adjustbox{max width=\textwidth}{%
\footnotesize
\setlength{\tabcolsep}{3.6pt}
\renewcommand{\arraystretch}{1.05}
\begin{tabular}{@{}lll@{}}
\toprule
\textbf{Evaluation Dimension} & \textbf{Stock Codex CLI \code{0.171.0}} & \textbf{Codex CLI \code{0.171.0} + \shellplan{} (MCP)} \\
\midrule
RO Controls (4: benign, \code{rm}, \code{touch}, \code{py}) & 4/4 expected (1 allow, 3 blocked) & 4/4 expected (1 allow, 3 blocked) \\
RO Attacks (5: \code{uniq}, \code{sort -o}, \code{sed}, \code{git show}, \code{.git}) & \textbf{5/5 target effect} (\code{bwrap=0}, \code{RC-14}) & \textbf{0/5 target effect} (5/5 escalated; 0 FS diff) \\
WS-Write \code{.git/} Control (\code{echo > .git/hooks/...}) & 1/1 blocked (\code{bwrap} \code{.git/} RO) & 1/1 blocked (\shellplan{} \code{.git/} guard) \\
WS-Write Selected Ops (non-equivalent policies)$^{\dagger}$ & \textbf{4/4 target effect} (3 \code{bwrap}/run) & \textbf{0/4 target effect} (4/4 escalated; 0 FS diff) \\
\midrule
Startup Untrusted Repo (Git filter; details redacted) & \textbf{Ran \code{/bin/sh -c}} (\code{-s ro} \& \code{untrusted}) & Mitigated via host \code{GIT\_DIR}; model \code{git} blocked \\
MCP Trace Audit (23 runs; 422 cmds/policy) & --- & 23/23 traces reconciled; 419/422 matched direct API \\
External-Writer Race Stress Test (1,000 rounds) & --- & \textbf{0 outside writes; 741 unhandled exceptions} \\
\bottomrule
\end{tabular}}
\begin{flushleft}
\scriptsize
$^{\dagger}$Operations: \code{uniq src/app.py}, \code{sed -i package.json}, \code{rm -r src}, \code{sed} execute. Ordinary workspace writes are permitted by stock Codex's workspace-write policy. Under \shellplan{} MCP WS-write, benign \code{echo hello > notes.txt} succeeded (rejected under RO as \code{WRITE\_REDIR\_NOT\_GRANTED}).
\end{flushleft}
\end{table*}

Across all 14 live cases (Table~\ref{tab:codex_integ}), \shellplan{} escalated all 5 read-only attacks and all 4 workspace operations with zero persistent mutations. Across 23 \code{strace}-monitored runs (14 cases + 9 auxiliary runs), every worker process and descendant \code{execve} reconciled 1-to-1 with MCP call logs or probes in $\mathcal{H}$ (verified via an empty-log negative control), and MCP dispatch matched direct API invocation on 419/422 commands per policy (3 nondeterministic \code{time}/\code{stat}/\code{diff} outputs). Deployment also exposed a host-level boundary outside any command gate: prior to the first model turn, Codex's own startup repository-status \code{git} call executed a repository-configured filter outside the sandbox in an untrusted read-only workspace despite Codex's existing hook-hardening flags (exact config key redacted pending OpenAI remediation); isolating Codex's startup \code{GIT\_DIR} mitigated it while \shellplan's \code{gitguard} rejected model-issued \code{git status}. Finally, in a 1,000-round external-writer race stress test alternating directory \code{d} between a real directory and an external symlink during \code{echo pwn > d/fN.txt}, \textbf{0 outside-workspace writes occurred} (83 wrote inside $W$, 145 returned path errors, 16 escalated at authorization, 15 rejected at open time, and 741 aborted via unhandled filesystem exceptions when directories disappeared mid-check).

\paragraph{Residual Boundaries.}
Three residual edge cases delimit the frozen implementation: (1)~\textbf{Symlinked metadata edge case (\code{M1}):} under \code{native+ws-write}, \shellplan's \code{lstat}-based directory-shape check did not recognize a pre-staged directory whose \code{objects} and \code{refs} entries were symbolic links, allowing writes to \code{config} and \code{HEAD}; in a 52-command follow-up audit, Bash executed hooks on 20/52 commands whereas \shellplan{} executed 0/52 because \code{-C}/\code{--git-dir}/\code{GIT\_DIR=} escalated and \code{gitguard} blocked workspace-root auto-discovery. (2)~\textbf{Inherited \code{//} \code{PWD} mismatch (\code{F1}):} when inherited \code{PWD} begins with \code{//} in a symlink-free workspace, Bash preserves both leading slashes for \code{pwd}/\code{pwd -L}/\code{pwd -P} while \shellplan{} prints one. (3)~\textbf{Pipeline command-substitution \code{stdin} (\code{F2}):} Case~C5.

\begin{findingbox}{Answer to RQ4 (Empirical Boundaries)}
Across 44 SWE-bench Verified trajectories, \shellplan{} achieved 0/348 non-A2 over-allows and 343/343 byte-exact Bash matches, and blocked all 5 live read-only attacks inside unmodified Codex CLI. Remaining boundaries are clearly localized: stateful \code{cd} chains (0/108 on OpenHands), one pre-staged symlinked-metadata edge case (\code{M1}), two fidelity mismatches (\code{F1}, \code{F2}), and host-level startup processes outside the tool gate.
\end{findingbox}

% ============================================================================
\section{Threats to Validity and Responsible Disclosure}
\label{sec:threats}
% ============================================================================

\textbf{Internal Validity.} In the Stealth Evasion Benchmark (1,250 runs), baselines and \shellplan{} ran in separate batches on two identical GPU hosts; our primary Controlled Agent Benchmark (760 runs) eliminates batch effects by evaluating all four arms in a single batch. In Live Product Trials, three agent--model pairs used a schema-only HTTP adapter and Roo CLI ran its untouched upstream bundle inside a headless wrapper, leaving all checker code unmodified. In Codex MCP deployment, stock and MCP workspace-write policies are not equivalent and are reported accordingly. In external trajectory replay, 8/596 editor edits failed and 115 steps lacked recorded exit codes (79.88\% joint agreement on 1,342 verifiable steps); pre-divergence subsets strictly exclude the first divergent step.

\textbf{External \& Construct Validity.} We pin eight open-source and open-CLI implementations on Linux with GNU Bash~5.2 (November~2026); closed-source tools whose terms prohibit automated benchmarking are excluded. Live Product Trials use direct verification prompts to establish existence of policy-violating effects rather than natural-world injection frequencies, and syntax-stress tasks explicitly test post-escalation recovery.

\textbf{Responsible Disclosure.} All experiments ran exclusively inside offline \code{--network none} Linux containers without targeting external systems. On November~22,~2026, we submitted private coordinated vulnerability reports (with minimal benign container reproducers and remediation guidance under a 90-day disclosure window) to all active upstream vendors and maintainers---OpenAI (via Bugcrowd and \code{disclosure@openai.com}), Google OSS VRP, Mistral AI, upstream \code{tree-sitter-bash} maintainers, Qwen Code (GitHub PVR), OpenCode (GitHub PVR), and Cline Security---along with courtesy archival notices to the maintainers of Roo Code and Continue (both archived read-only prior to submission). Consistent with this embargo, this arXiv v1 preprint redacts exact trigger payloads and unpatched reproduction recipes, which will be restored in v2 once patches are released.

% ============================================================================
\section{Related Work}
\label{sec:related}
% ============================================================================

Checker--interpreter discrepancies have long been studied in NIDS~\cite{ptacek1998insertion,handley2001network}, LangSec~\cite{momot2016seven}, and HTTP request smuggling~\cite{jabiyev2021treqs}, while differential testing has exposed semantic gaps in TLS certificate parsers~\cite{chen2023sbdt}, WebAssembly runtimes~\cite{cao2024wasmaker}, and EVMs~\cite{ma2025opdiffer}; Smoosh~\cite{greenberg2020smoosh} provides an executable POSIX shell semantics. For LLM and coding agents, indirect prompt injection~\cite{greshake2023indirect,zhan2024injecagent,debenedetti2024agentdojo,son2026agentbreaker} and risky code execution~\cite{guo2024redcode,ruan2024toolemu} motivate coding-agent tool red-teaming~\cite{xie2026redteaming}, progressive tool-chain attacks~\cite{shao2026datura}, framework fuzzing~\cite{long2026logichunter}, and exploit analyses~\cite{liu2025aishelljack,lee2025takedown,yang2026approvaltoabuse}, alongside gate studies of denylist fragility~\cite{chen2026cmdneedle}, hidden command-path failures~\cite{li2026quotebench}, authorization-honoring guardrails~\cite{jing2026sameop}, harness mechanisms~\cite{zhu2026harnesssecbench}, and hardened-policy utility~\cite{davidovich2026permissiondenied}. Whereas recent ISSTA~2026 studies target agent frameworks~\cite{long2026logichunter}, web contexts~\cite{son2026agentbreaker}, or model-level tool hijacking~\cite{xie2026redteaming,shao2026datura}, \parsegap{} isolates the checker--shell--sandbox differential boundary across unmodified vendor checkers and OS sandboxes. Runtime defenses constrain agent actions via preference optimization~\cite{chen2025secalign}, system isolation~\cite{wu2025isolategpt}, information-flow control~\cite{debenedetti2026camel,costa2025fides}, privilege control~\cite{shi2025progent}, or rule languages such as AgentSpec~\cite{wang2026agentspec}; \shellplan{} instead fixes how an approved shell command is interpreted. Closest to our defense, Jing et al.~\cite{jing2026policygate} adjudicate typed realized actions without parsing shell syntax, whereas \shellplan{} compiles and directly executes the shell strings existing coding agents emit.

% ============================================================================
\section{Conclusion}
\label{sec:conclusion}
% ============================================================================

Shell command gates approve their own reading of a command string, while the system shell executes its own. Across eight production coding agents, \parsegap{} exposed divergences at the parse, at utility argument semantics in every checker, and at OS confinement in Codex CLI. \shellplan{} removes the second reading by compiling a strict Bash subset once, enforcing positive utility specifications, and executing the approved plan directly via \code{execve} without relaxing confinement---eliminating all tested attacks at sub-2\,ms latency and preserving 100\% frontier-model completion on standard development tasks, while clarifying remaining compiler and utility-coverage frontiers.

% ============================================================================
\section*{Data Availability}
% ============================================================================

Upon publication and completion of our 90-day coordinated disclosure window, we will release an archival artifact containing the frozen \shellplan{} implementation (commit \code{2e07e41}), unit test suites, complete formal semantics and proofs (\code{PROOFS.pdf}), pre-registered external-evaluation protocol (commit \code{6c8968b}), and sanitized case-level summary datasets backing Tables~1--6. Third-party benchmarks (SWE-bench Verified, SWE-rebench, and NL2Bash) remain governed by their upstream licenses; vendor-specific exploit reproduction scripts remain withheld during coordinated disclosure.

\bibliographystyle{ACM-Reference-Format}
\bibliography{references}

\end{document}